\documentclass[preprint,showpacs,preprintnumbers,amsmath,amssymb,floatfix,endfloats*]{revtex4}

\usepackage{graphicx}
\usepackage{dcolumn}
\usepackage{bm}
\usepackage{amsfonts}

\DeclareMathAlphabet{\mathsfsl}{OT1}{cmr}{bx}{it}
\begin{document}
\title{Interfacial friction between semiflexible polymers and crystalline surfaces}
\author{Nikolai V. Priezjev}
\affiliation{Department of Mechanical Engineering, Michigan State
University, East Lansing, Michigan 48824}
\date{\today}
%
\begin{abstract}

The results obtained from molecular dynamics simulations of the
friction at an interface between polymer melts and weakly attractive
crystalline surfaces are reported.  We consider a coarse-grained
bead-spring model of linear chains with adjustable intrinsic
stiffness.  The structure and relaxation dynamics of polymer chains
near interfaces are quantified by the radius of gyration and decay
of the time autocorrelation function of the first normal mode.  We
found that the friction coefficient at small slip velocities
exhibits a distinct maximum which appears due to shear-induced
alignment of semiflexible chain segments in contact with solid
walls.  At large slip velocities the decay of the friction
coefficient is independent of the chain stiffness.  The data for the
friction coefficient and shear viscosity are used to elucidate main
trends in the nonlinear shear rate dependence of the slip length.
The influence of chain stiffness on the relationship between the
friction coefficient and the structure factor in the first fluid
layer is discussed.

\end{abstract}



\maketitle

\section{Introduction}

Understanding the interfacial rheology of complex fluids is
important in many processes relevant to technological applications
including polymer processing~\cite{Denn01,Achilleos02}, boundary
lubrication~\cite{Marina10}, and dewetting of polymer
films~\cite{JacobsRev10}.  Numerous experimental studies have
demonstrated that flow velocity profiles in nanoconfined systems can
be significantly influenced by slip at polymer-solid
interfaces~\cite{Wang99,Bhushan08}.  The measure of slippage is the
so-called slip length, which is defined as the distance between the
physical interface and imaginary plane where the extrapolated
velocity profile reaches the substrate velocity.   During the last
two decades, the dependence of the slip length on flow conditions
and material properties of substrates was extensively investigated
by molecular dynamics (MD) simulations for
monatomic~\cite{Fischer89,KB89,Thompson90,Nature97,Barrat99,Barrat99fd,Attard04,Priezjev07,PriezjevJCP,Kovalenko08,Li09,Yong10,Asproulis10,Niavarani10,Asproulis11,Freund11,WangZhao11,Kannam11,Priezjev11,Kannam12}
and
polymeric~\cite{Toxvaerd94,Thompson95,Manias96,dePablo96,StevensJCP97,Koike98,Tanner99,Tanner02,Priezjev04,Priezjev08,Servantie08,LichPRL08,Priezjev09,Hendy09,Denniston10,Priezjev10,Lichter11,Farrow11,Zaccheddu12}
fluids.

\begin{figure}[t]
\vspace*{-3mm}
\includegraphics[width=8.0cm,angle=0]{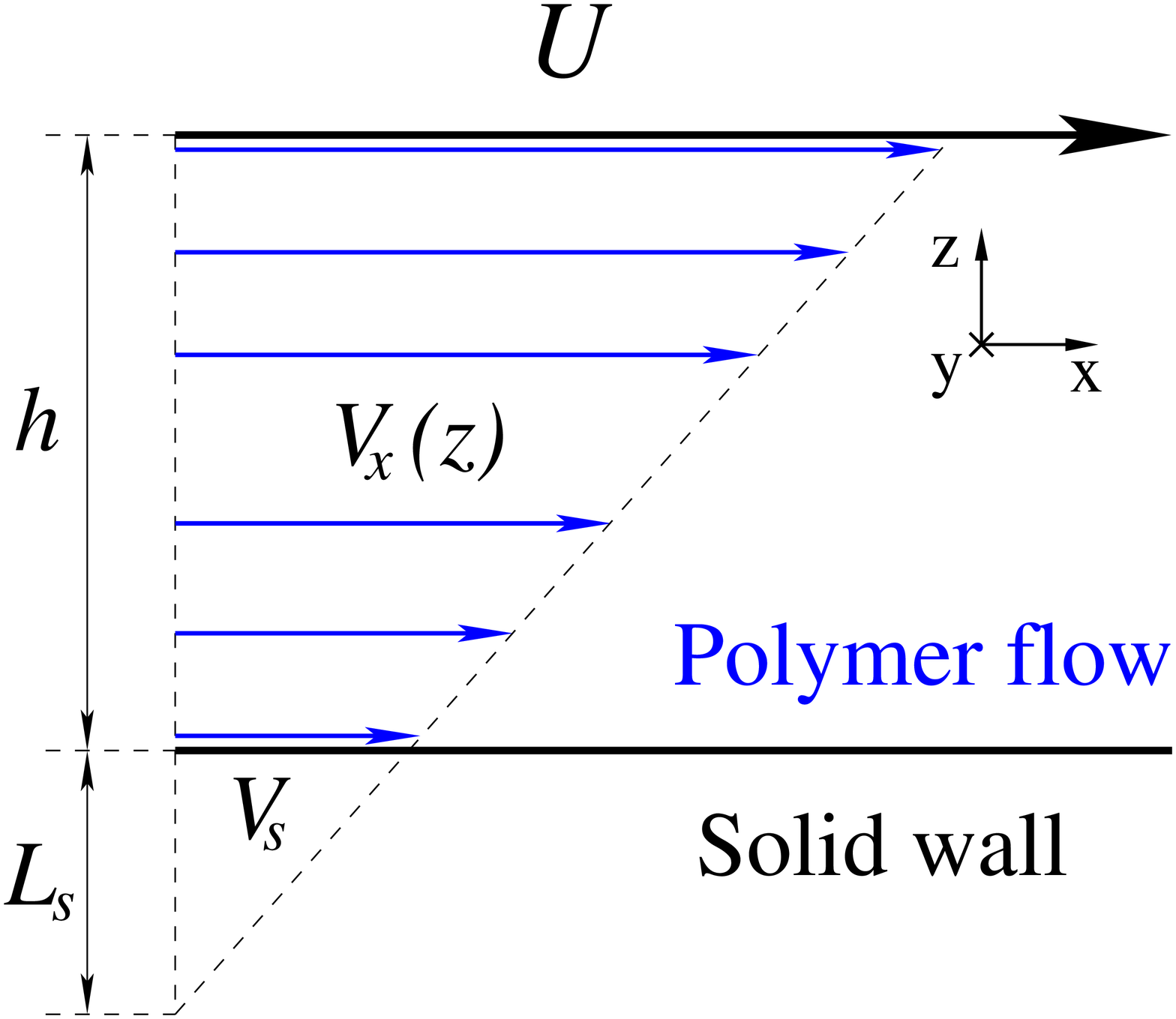}
\caption{(Color online) A schematic view of the Couette flow
configuration with slip at the lower and upper walls. Steady shear
flow is generated by the upper wall moving with a constant velocity
$U$ in the $\hat{x}$ direction while the lower wall is at rest. The
slip length and the slip velocity are related via
$V_s=\dot{\gamma}L_s$, where $\dot{\gamma}$ is the shear rate
computed from the slope of the velocity profile. }
\label{Fig:schematic}
\end{figure}

In the case of simple shear flow illustrated schematically in
Fig.\,\ref{Fig:schematic}, the shear stress is the same in the bulk
of the channel and at the interface; and, therefore, the slip length
can be calculated from the ratio of the fluid viscosity to the
friction coefficient at the interface according to
\begin{equation}
L_s=\frac{\mu}{k}, \label{Eq:ls_fr_k}
\end{equation}
where the friction coefficient $k$ is defined by the relation
between the slip velocity and wall shear stress~\cite{deGennes94}.
At equilibrium, both the fluid viscosity~\cite{Evans80,Bellemans87}
and friction
coefficient~\cite{Barrat99fd,ServantieJCP08,Kovalenko08,Kannam11}
can be estimated from the Green-Kubo relations, and the slip length
in the limit of zero shear rate is then computed from
Eq.\,(\ref{Eq:ls_fr_k}).    In the presence of flow, the fluid
viscosity and friction coefficient might depend on shear rate and
slip velocity respectively; and, as a result, the slip length is
often a nonlinear function of shear
rate~\cite{Nature97,Priezjev07,Priezjev08,LichPRL08,Priezjev10,Freund11,WangZhao11,Kannam11}.
For example, it was recently shown that at an interface between
unentangled polymer melts and passive surfaces (no chemical bonds
with the surface and weak wall-fluid interaction energy), the slip
length passes through a local minimum at low shear rates and then
increases rapidly at higher shear
rates~\cite{Priezjev08,Priezjev10}.   This non-monotonic behavior
was explained by computing the rate-dependent viscosity and the
friction coefficient that undergoes a transition from a constant
value to the power-law decay as a function of the slip
velocity~\cite{Priezjev08,Priezjev10}.   One of the motivations of
the present study is to examine whether these conclusions hold for
different simulation ensembles and intramolecular potentials.

In the past two decades, a number of MD studies have demonstrated
that the degree of slip at the interface between molecular liquids
and crystalline surfaces depends on the structure of the first fluid
layer in contact with the periodic surface
potential~\cite{Thompson90,Barrat99fd,Attard04,Priezjev07,PriezjevJCP,Kovalenko08,Priezjev08,Priezjev09,Priezjev10}.
In general, the interfacial slip is suppressed with increasing
height of the peak in the fluid structure factor computed at the
main reciprocal lattice vector.   In turn, the in-plane order within
the adjacent fluid layer is determined by several factors, including
the commensurability of the liquid and solid
structures~\cite{Thompson90,Barrat99fd,Attard04,Priezjev07,Priezjev10},
wall-fluid interaction
energy~\cite{Thompson90,Barrat99fd,Priezjev07,Yong10,Priezjev10},
surface
rigidity~\cite{Thompson90,PriezjevJCP,Asproulis10,Priezjev10},
molecular
structure~\cite{Thompson95,Priezjev04,Kovalenko08,Priezjev10,Lichter11},
and fluid pressure~\cite{Thompson95,Barrat99,Priezjev09,Priezjev10}.
Interestingly, recent simulation results have shown that the
friction coefficient in the linear slip regime is a function of a
combined variable that is a product of the height of the main peak
in the structure factor and the contact density of the first fluid
layer near the solid wall~\cite{Priezjev10}.  However, at present,
there exists no exact relationship between the friction coefficient
(or the slip length) and the microscopic properties of the
liquid-solid interface.

The formation of the interfacial fluid layer and hydrodynamic
boundary conditions can be significantly affected by the polymer
chain architecture.   For instance, several MD studies have reported
flow profiles with a finite slip velocity in thin alkane films where
stiff polymer chains tend to align in layers parallel to the
surface~\cite{Toxvaerd94,StevensJCP97,Koike98,Tanner99}.   On the
other hand, a weaker density layering near the wall and pronounced
slippage were observed for branched alkane
molecules~\cite{Tanner02}.  Notably, it was demonstrated that upon
increasing the length of linear, freely-jointed chains, the
structure of the first fluid layer near a solid wall is reduced,
resulting in smaller values of the friction coefficient and larger
slip lengths at low shear rates~\cite{Priezjev04}.   Later, it was
found that the existence and location of a double bond along the
backbone of linear oligomers affect the structure of the interfacial
fluid layer near crystalline aluminum walls and lead to either
negative or large positive slip lengths~\cite{Denniston10}.   More
recently, it was shown that slip velocity is reduced for liquids
which consist of molecules that can easily conform their atoms into
low-energy sites of the substrate potential~\cite{Lichter11}.
Despite extensive research on the fluid structure and shear response
in thin polymer films, it is often difficult to predict even
qualitatively the influence of liquid molecular structure on the
interfacial slip.

In this paper, we investigate the effect of chain bending stiffness
on the fluid structure and friction coefficient at interfaces
between linear polymers and crystalline surfaces.  We will show that
the relaxation dynamics near weakly attractive surfaces is
significantly slowed down for stiffer chains at equilibrium.  The
orientation of semiflexible chains in shear flow leads to the
enhanced density layering away from the walls and partial alignment
of extended chain segments in the first fluid layer.  It will be
demonstrated that the shear-induced ordering of the chain segments
produces a distinct maximum in the liquid structure factor and the
friction coefficient at small slip velocities. Finally, the
characteristic features in the shear rate dependence of the slip
length are interpreted in terms of the shear-thinning viscosity and
the dynamic friction coefficient.

The rest of this paper is organized as follows. The details of
molecular dynamics simulations, interaction potentials, and
equilibration procedure are described in Section~\ref{sec:MD_Model}.
The simulation results are presented in Section~\ref{sec:Results}.
More specifically, the chain conformation and relaxation dynamics
are analyzed in~\ref{subsec:conformation_relaxation},  examples of
density and velocity profiles are presented
in~\ref{subsec:density_velocity}, shear viscosity and slip lengths
are reported in~\ref{subsec:viscosity_slip_length}, the velocity
dependence of the friction coefficient is examined
in~\ref{subsec:friction_velocity}, and, finally, the fluid structure
near solid walls is considered in~\ref{subsec:friction_structure}.
Brief conclusions are given in the last section.

\section{Molecular dynamics simulation model}
\label{sec:MD_Model}

We consider a coarse-grained bead-spring model of unentangled
polymer melt, which consists of $M=480$ linear chains of $N=20$
beads (or monomers) each. In this model any two fluid monomers
interact via the truncated Lennard-Jones (LJ) potential
\begin{equation}
V_{LJ}(r)=4\,\varepsilon\,\Big[\Big(\frac{\sigma}{r}\Big)^{12}\!-\Big(\frac{\sigma}{r}\Big)^{6}\,\Big],
\label{Eq:LJ}
\end{equation}
where $\varepsilon$ and $\sigma$ are the energy and length scales of
the fluid phase. The cutoff radius $r_c=2.5\,\sigma$ and the total
number of fluid monomers $N_{f}=9600$ are fixed throughout all
simulations. Similarly, fluid monomers interact with wall atoms via
the LJ potential with the following parameters $\varepsilon_{\rm
wf}=0.8\,\varepsilon$, $\sigma_{\rm wf}=\sigma$, and
$r_c=2.5\,\sigma$.

\begin{figure}[t]
\includegraphics[width=13.0cm,angle=0]{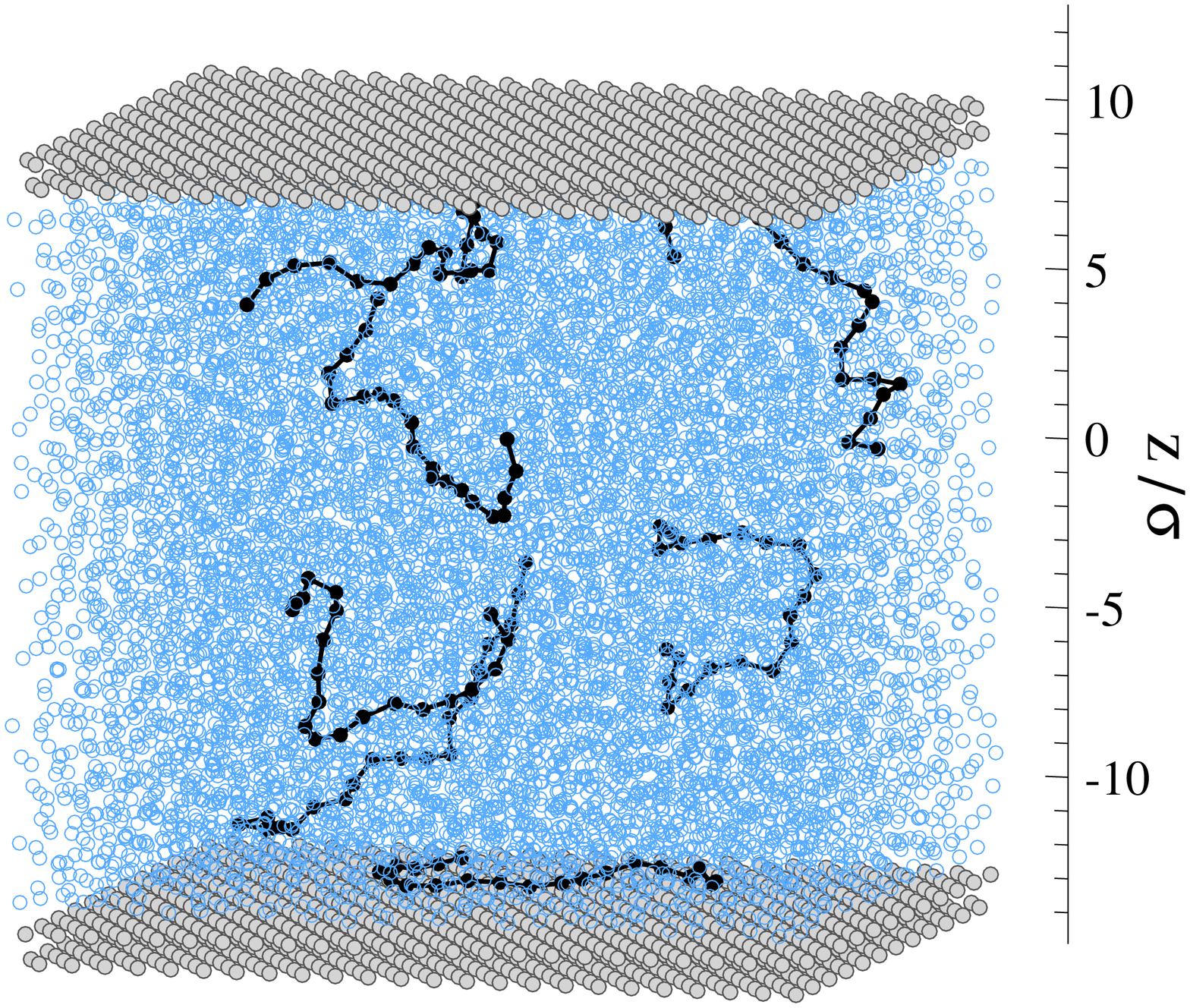}
\caption{(Color online) Instantaneous positions of fluid monomers
(open blue circles) and fcc wall atoms (filled gray circles) at
equilibrium (i.e., both walls are at rest). Each monomer belongs to
a polymer chain ($N=20$) with the bending stiffness coefficient
$k_{\theta}=2.5\,\varepsilon$. Seven chains are indicated by thick
solid lines and filled black circles. The fluid monomer density is
$\rho=0.91\,\sigma^{-3}$ and the wall atom density is
$\rho_w=1.40\,\sigma^{-3}$.}
\label{Fig:snapshot}
\end{figure}

Any two consecutive beads in a polymer chain interact through the
finitely extensible nonlinear elastic (FENE) potential~\cite{Bird87}
\begin{equation}
V_{FENE}(r)=-\frac{k_s}{2}\,r_{\!o}^2\ln[1-r^2/r_{\!o}^2],
\label{Eq:FENE}
\end{equation}
with the standard parameters $k_s=30\,\varepsilon\sigma^{-2}$ and
$r_{\!o}=1.5\,\sigma$~\cite{Kremer90}. The combination of LJ and
FENE potentials yields an effective bond potential between the
nearest-neighbor beads with the average bond length
$b=0.97\,\sigma$~\cite{Kremer90}. This bond potential is strong
enough to prevent chain crossing and breaking even at the highest
shear rates considered in the present study.  In addition, the
flexibility of polymer chains is controlled by the bending potential
as follows:
\begin{equation}
U_{bend}(\theta) = k_{\theta}\,(1-\textrm{cos}\,\theta),
\label{Eq:U_bend}
\end{equation}
where $k_{\theta}$ is the bending stiffness coefficient and $\theta$
is the angle between two consecutive bonds along a polymer
chain~\cite{GrestBend03}. In the present study, the bending
stiffness coefficient was varied in the range $0\leqslant
k_{\theta}\leqslant 3.5\,\varepsilon$. A snapshot of the confined
polymer melt that consists of semiflexible linear chains with the
bending coefficient $k_{\theta}=2.5\,\varepsilon$ is shown in
Figure\,\ref{Fig:snapshot}.

In order to remove viscous heating generated in the shear flow, the
motion of fluid monomers was coupled to an external heat bath via a
Langevin thermostat~\cite{Grest86} applied in the $\hat{y}$
direction to avoid bias in the shear flow direction (the $\hat{x}$
direction). This is a standard thermostatting procedure often used
in MD simulations of sheared
fluids~\cite{Thompson90,Thompson95,Nature97,Priezjev05,Priezjev06}.
Thus, the equations of motion for fluid monomers are summarized as
follows:
\begin{eqnarray}
\label{Eq:Langevin_x}
m\ddot{x}_i & = & -\sum_{i \neq j} \frac{\partial V_{ij}}{\partial x_i}\,, \\
\label{Eq:Langevin_y}
m\ddot{y}_i + m\Gamma\dot{y}_i & = & -\sum_{i \neq j} \frac{\partial V_{ij}}{\partial y_i} + f_i\,, \\
\label{Eq:Langevin_z}
m\ddot{z}_i & = & -\sum_{i \neq j} \frac{\partial V_{ij}}{\partial z_i}\,, %
\end{eqnarray}
where $\Gamma=1.0\,\tau^{-1}$ is the friction coefficient that
controls the damping term, $V_{ij}$ is the total interaction
potential, and $f_i$ is a random force with zero mean and variance
$\langle f_i(0)f_j(t)\rangle=2mk_BT\Gamma\delta(t)\delta_{ij}$
obtained from the fluctuation-dissipation theorem.  The Langevin
thermostat temperature is set $T=1.1\,\varepsilon/k_B$, where $k_B$
refers to the Boltzmann constant.  The equations of motion were
solved numerically using the fifth-order Gear predictor-corrector
algorithm~\cite{Allen87} with a time step $\triangle t=0.005\,\tau$,
where $\tau=\sqrt{m\sigma^2/\varepsilon}$ is the LJ time.  Typical
values of the length, energy, and time scales for hydrocarbon chains
are $\sigma=0.5\,\text{nm}$, $\varepsilon=30\,\text{meV}$, and
$\tau=3\times10^{-12}\,\text{s}$~\cite{Allen87}.


The polymer melt is confined between two crystalline walls as
illustrated in Figure\,\ref{Fig:snapshot}.  Each wall consists of
$1152$ atoms distributed between two layers of the face-centered
cubic (fcc) lattice with density $\rho_w=1.40\,\sigma^{-3}$. For
computational efficiency, the wall atoms are fixed rigidly to the
wall lattice sites, which form two $(111)$ planes with $[11\bar{2}]$
orientation parallel to the $\hat{x}$ direction.  The
nearest-neighbor distance between the lattice sites within the
$(111)$ plane is $d=1.0\,\sigma$ and the first reciprocal lattice
vector in the $\hat{x}$ direction is
$\mathbf{G}_{1}=(7.23\,\sigma^{-1},0)$.  The channel dimensions in
the $xy$ plane are measured to be $L_x=20.86\,\sigma$ and
$L_y=24.08\,\sigma$.  Periodic boundary conditions were applied
along the the $\hat{x}$ and $\hat{y}$ directions parallel to the
solid walls.

In our simulations, the distance between the wall lattice planes,
which are in contact with the fluid phase, was fixed at
$h=22.02\,\sigma$.  Hence, the volume accessible to the fluid phase
corresponds to the fluid monomer density
$\rho=N_{f}/L_{x}L_{y}(h-\sigma)=0.91\,\sigma^{-3}$; and, in the
absence of shear flow, the resulting fluid pressure and temperature
are $1.0\,\varepsilon\,\sigma^{-3}$ and $1.1\,\varepsilon/k_B$
respectively.  In the present study, the relatively low polymer
density (or normal pressure) was chosen based on the results from
our previous study where it was shown that for weak wall-fluid
interactions and $\rho\leqslant 1.02\,\sigma^{-3}$ (or $P\leqslant
5.0\,\varepsilon\,\sigma^{-3}$), the fluid velocity profiles remain
linear in a wide range of shear rates~\cite{Priezjev08}.  In
contrast, it was demonstrated that at higher polymer densities (or
pressures), the velocity profiles acquire a pronounced curvature
near the wall and the relaxation of flexible polymer chains in the
interfacial region becomes very slow~\cite{Priezjev09}.


The system was first equilibrated for about $5\times10^4\tau$ while
both walls were at rest. Then, the velocity of the upper wall was
increased gradually up to a target value, followed by an additional
equilibration period of about $5\times10^4\tau$. In this study, the
upper wall velocity was varied over about three orders of magnitude
$0.005\leqslant U\,\tau/\sigma \leqslant 5.5$. Once the steady shear
flow was generated, the velocity, density, and temperature profiles
were averaged within horizontal bins of thickness $\Delta
z=0.01\,\sigma$ for a time period up to $5\times10^5\tau$. At the
lowest upper wall speed, $U=0.005\,\sigma/\tau$, the velocity
profiles were computed in $24$ independent systems for the total
time period of about $5\times10^6\tau$. An upper estimate of the
Reynolds number at high shear rates is $Re=\rho\,\!h\,U/\mu=O(10)$,
which is indicative of laminar flow conditions in the channel.

We finally note that MD simulations were also performed at a
constant normal load, where the distance between the walls was
allowed to vary under the constant normal pressure
$P_{\perp}=1.0\,\varepsilon\,\sigma^{-3}$ applied to the upper wall.
However, we did not observe any qualitatively new behavior; and for
the sake of brevity, these results are not reported in the present
study.

\section{Results}
\label{sec:Results}

\subsection{Chain conformation and relaxation dynamics}
\label{subsec:conformation_relaxation}

\begin{figure}[t]
\includegraphics[width=12.cm,angle=0]{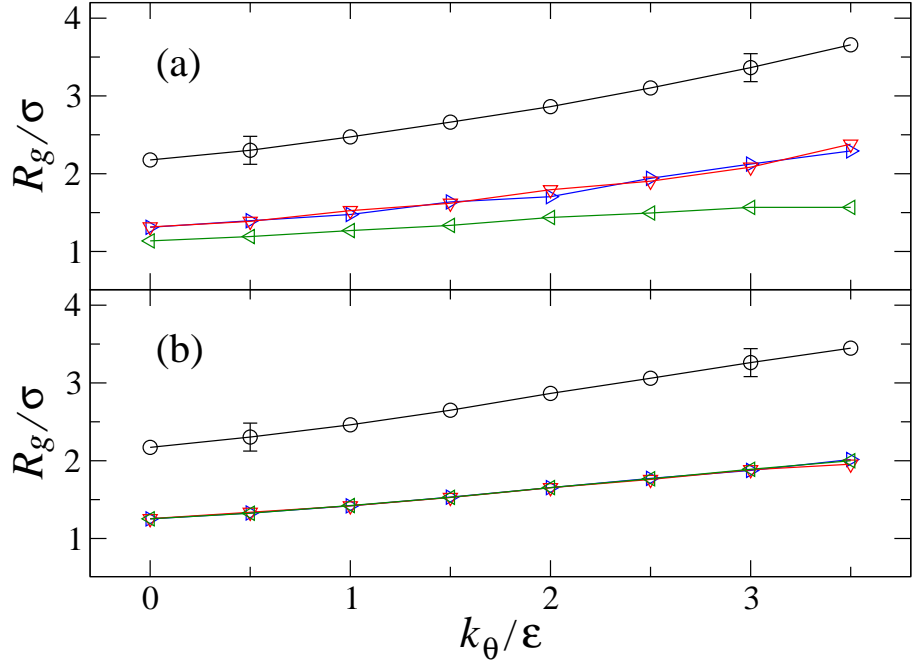}
\caption{(Color online) The averaged $\hat{x}$, $\hat{y}$, and
$\hat{z}$ components of the radius of gyration $R_{gx}$
($\triangleright$), $R_{gy}$ ($\triangledown$), $R_{gz}$
($\triangleleft$), and the total radius of gyration $R_{g}$
($\circ$) as a function of $k_{\theta}$ for polymer chains (a) in
contact with the solid walls and (b) in the bulk region. The
simulations were performed at the constant fluid density
$\rho=0.91\,\sigma^{-3}$ while both walls were at rest.}
\label{Fig:Rg_bulk_walls}
\end{figure}

The spatial configuration of polymer chains is well characterized by
the radius of gyration, which is defined as the average distance
between monomers in a polymer chain and its center of mass as
follows:
\begin{equation}
R_g^2=\frac{1}{N}\sum_{i=1}^{N}(\mathbf{r}_i-\mathbf{r}_{\text{cm}})^2,
\label{Eq:Rg_def_eq}
\end{equation}
where $\mathbf{r}_i$ is the position vector of the $i$-th monomer,
$N=20$ is the number of monomers per chain, and
$\mathbf{r}_{\text{cm}}$ is the chain center of mass defined as
\begin{equation}
\mathbf{r}_{\text{cm}}=\frac{1}{N}\sum_{i=1}^{N}\mathbf{r}_i.
\label{Eq:Rcm_def_eq}
\end{equation}
Figure \ref{Fig:Rg_bulk_walls} shows the radius of gyration and its
components along the $\hat{x}$, $\hat{y}$, and $\hat{z}$ directions
as a function of the bending stiffness coefficient.   The chain
statistics were collected in the interfacial regions (where at least
one monomer in a chain is in contact with wall atoms) and in the
middle of the channel (the chain center of mass is located further
than $6\,\sigma$ away from the walls).   As expected, in both cases
$R_g$ increases with increasing chain stiffness. Note that even at
the largest value of the bending stiffness coefficient
$k_{\theta}=3.5\,\varepsilon$, the size of polymer chains is smaller
than the channel dimensions.   The simulation results in
Fig.\,\ref{Fig:Rg_bulk_walls} indicate that in the bulk region the
chain configuration is isotropic, while near the interfaces polymer
chains become flattened, i.e., $R_{gz}<R_{gx}\approx R_{gy}$, which
is in agreement with previous MD
studies~\cite{Bitsanis90,Doi01,Niavarani08}. It is also apparent
that fully flexible chains in contact with the walls are packed on
average within the first two fluid layers [\,$2R_{gz}\approx
2\,\sigma$ for $k_{\theta}=0$ in
Fig.\,\ref{Fig:Rg_bulk_walls}\,(a)\,].  In contrast, semiflexible
chains extend up to about three molecular diameters from the walls.
Visual inspection of the polymer chains in the interfacial regions
revealed that the conformation of semiflexible chains consists of
locally extended segments within the first fluid layer and segments
of several monomers oriented away from the walls.   Finally,
regardless of the chain stiffness, the total radius of gyration is
nearly the same in the bulk and close to the walls due to the
relatively weak wall-fluid interaction energy.

The local relaxation dynamics in confined polymer films can be
described by the decay of the time autocorrelation function of
normal modes~\cite{Bitsanis93,ManiasTh95,Doi01,Priezjev09}.  By
definition, the normal coordinates for a polymer chain that consists
of $N$ monomers are given by
\begin{equation}
\mathbf{X}_p(t)=\frac{1}{N}\sum_{i=1}^N
\mathbf{r}_i(t)\,\textrm{cos}\,\frac{p\pi(i-1)}{N-1},
\label{Eq:normal_mode_eq}
\end{equation}
where $\mathbf{r}_i$ is the position vector of the $i$-th monomer in
the chain, and $p=0, 1,..., N-1$ is the mode number~\cite{Binder95}.
The longest relaxation time corresponds to the first mode $p=1$,
i.e., to the relaxation of the whole chain~\cite{Binder95}.  The
normalized time autocorrelation function for the first normal mode
is then defined as follows:
\begin{equation}
C_1(t)=\langle \mathbf{X}_1(t)\cdot \mathbf{X}_1(0)\rangle/ \langle
\mathbf{X}_1(0)\cdot \mathbf{X}_1(0)\rangle. \label{Eq:auto_corr_eq}
\end{equation}
In our study, the autocorrelation function
[Eq.\,(\ref{Eq:auto_corr_eq})] was computed separately in the
interfacial regions (where the chain center of mass is confined
within $3\,\sigma$ from the walls) and in the bulk of the channel
where the fluid density is uniform (the center of mass is located at
least $6\,\sigma$ away from the walls).  An important aspect is that
the autocorrelation function was averaged only for those polymer
chains whose centers of mass remained within either the interfacial
or bulk regions during the relaxation time interval.

\begin{figure}[t]
\includegraphics[width=12.cm,angle=0]{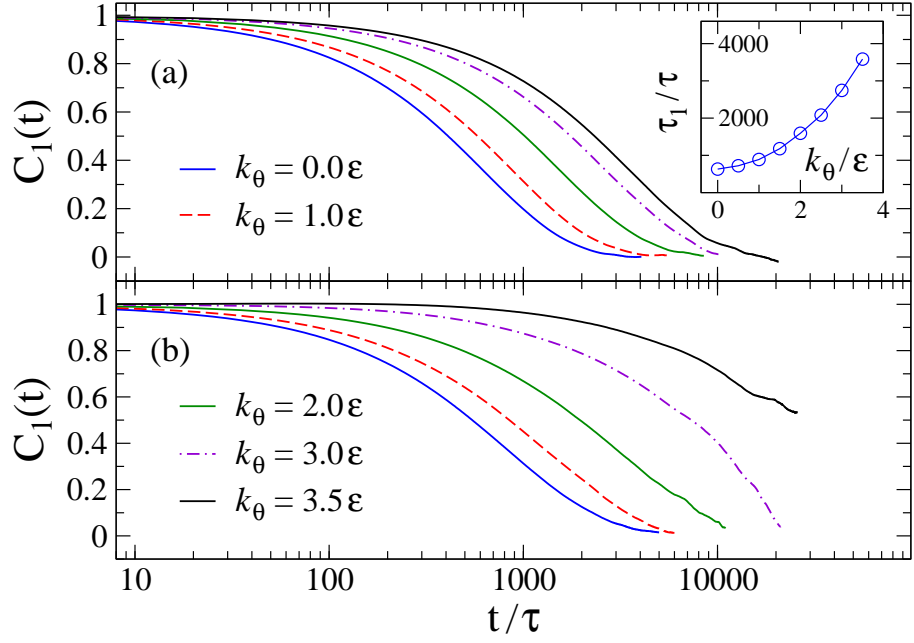}
\caption{(Color online) The time autocorrelation function of the
first normal mode Eq.\,(\ref{Eq:auto_corr_eq}) for polymer chains
(a) in the bulk region and (b) near the walls for several values of
the bending stiffness coefficient.  The inset shows the relaxation
time of polymer chains in the bulk.}
\label{Fig:auto_bulk_walls}
\end{figure}

Figure\,\ref{Fig:auto_bulk_walls} shows the relaxation of the time
autocorrelation function at equilibrium (i.e., when both walls are
at rest) for selected values of the bending stiffness coefficient.
As is evident from Fig.\,\ref{Fig:auto_bulk_walls}\,(a), the
relaxation rate of polymer chains in the bulk region decreases with
increasing bending stiffness.  A similar effect was reported
previously for linear bead-spring chains with variable bending
rigidity~\cite{Giessen05}, indicating that the reorientation
dynamics in the melt is slowed down for more rigid polymer chains.
In our study, the decay rate of the autocorrelation function in the
bulk is well described by the exponential function
$C_1(t)=\text{exp}(-t/\tau_{1})$, where $\tau_{1}$ is the
characteristic relaxation time. The inset in
Fig.\,\ref{Fig:auto_bulk_walls}\,(a) presents the variation of
$\tau_{1}$ as a function of the bending stiffness coefficient. The
inverse relaxation time, $1/\tau_{1}$, is related to the
characteristic shear rate, above which the shear viscosity is
expected to exhibit non-Newtonian behavior (see discussion below).
We note that for stiffer chains, the estimated shear rate is about
$2.5\times10^{-4}\tau^{-1}$, which is about the lowest shear rate
accessible in coarse-grained MD simulations (without excessive
computational time requirements).


In contrast, the decay in time of the autocorrelation function is
much slower for semiflexible polymer chains in the interfacial
regions, see Fig.\,\ref{Fig:auto_bulk_walls}\,(b).  Similar results
were observed previously in MD simulations of freely-jointed
$5$-mers adsorbed on weakly physisorbing surfaces, i.e., the
relaxation time of adsorbed chains is about an order of magnitude
larger than in the bulk~\cite{Bitsanis93}.  In our setup, the
relaxation time of flexible chains in the interfacial region is only
slightly larger than in the bulk, which means that the relaxation
dynamics is weakly affected by the substrate.  With increasing chain
stiffness, however, the rotational relaxation is significantly
slowed down. The data shown in Fig.\,\ref{Fig:auto_bulk_walls}\,(b)
cannot be well fitted by the single exponential function.  We
comment that the typical relaxation time for polymer chains in the
interfacial regions was used to determine the time interval for
averaging the radius of gyration and bond orientation.  It should
also be mentioned that test simulations of polymer chains with
larger stiffness coefficients, $k_{\theta}=4.0\,\varepsilon$ and
$4.5\,\varepsilon$, have shown that their relaxation dynamics near
interfaces is extremely slow and cannot be accurately resolved
(results not reported).


\subsection{Fluid density and velocity profiles}
\label{subsec:density_velocity}

\begin{figure}[t]
\includegraphics[width=12.cm,angle=0]{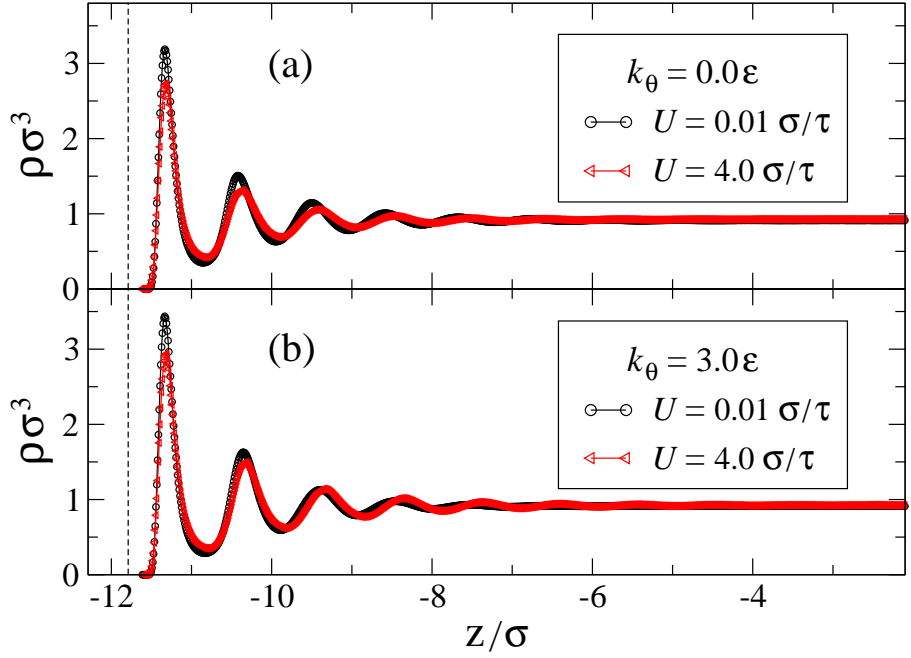}
\caption{(Color online) Averaged monomer density profiles near the
lower stationary wall for the indicated values of the upper wall
velocity $U$ and $\rho=0.91\,\sigma^{-3}$. The bending stiffness
coefficients are (a) $k_{\theta}=0.0\,\varepsilon$ and (b)
$k_{\theta}=3.0\,\varepsilon$. The left vertical axis at
$z=-12.29\,\sigma$ coincides with the fcc lattice plane in contact
with the polymer melt. The vertical dashed line at
$z=-11.79\,\sigma$ denotes the location of the liquid-solid
interface. }
\label{Fig:mol_dens_k}
\end{figure}

The averaged monomer density profiles are presented in
Fig.\,\ref{Fig:mol_dens_k} for fully flexible
($k_{\theta}=0.0\,\varepsilon$) and semiflexible
($k_{\theta}=3.0\,\varepsilon$) polymer chains at small
($U=0.01\,\sigma/\tau$) and large ($U=4.0\,\sigma/\tau$) upper wall
velocities.  Near the solid walls, these profiles exhibit typical
density oscillations that gradually decay to a uniform profile in
the middle of the channel.  Notice that the magnitude of the first
peak in the density profiles (defined as the contact density) is
higher for stiffer polymer chains.  For example, the contact density
is $\rho_c=3.19\,\sigma^{-3}$ for flexible chains and
$\rho_c=3.44\,\sigma^{-3}$ for $k_{\theta}=3.0\,\varepsilon$ when
the upper wall velocity is $U=0.01\,\sigma/\tau$ in
Fig.\,\ref{Fig:mol_dens_k}. This result, at first glance, appears to
be somewhat counterintuitive because one might expect that flexible
chains can pack more effectively near a flat surface.  However, with
increasing bending rigidity, the persistence length of polymer
chains increases; and, therefore, the first fluid layer contains
more extended chain segments.  When $U=0.01\,\sigma/\tau$ in
Fig.\,\ref{Fig:mol_dens_k}, the average number of consecutive
monomers per polymer chain in the first fluid layer is
$N_{seg}\approx3.1$ for flexible chains and $N_{seg}\approx5.2$ for
$k_{\theta}=3.0\,\varepsilon$. It turns out that these locally
extended chain segments arrange themselves more tightly near the
surface.  We also note that similar trends in the fluid density
layering were reported in other coarse-grained MD simulations;
namely, that with increasing length of (semi)flexible polymer
chains, the amplitude of density oscillations near a solid wall
becomes (larger) smaller~\cite{Koike98,Bitsanis90}.

As shown in Fig.\,\ref{Fig:mol_dens_k}\,(a), the height of the
density peaks in the case of flexible chains is reduced at the
higher upper wall speed $U=4.0\,\sigma/\tau$.  At these flow
conditions, the slip velocity of the first fluid layer is relatively
large (of about $1.0\,\sigma/\tau$), and the temperature of the
fluid near the walls is higher than the temperature of the Langevin
thermostat, leading to a reduced density layering.  This is
consistent with the results of previous MD studies where the shear
response of thin polymer films was examined in a wide range of shear
rates~\cite{Priezjev08,Priezjev09,Priezjev10}.  Interestingly, while
the amplitude of the first two peaks in the density profile for
semiflexible chains ($k_{\theta}=3.0\,\varepsilon$) is also reduced
at the higher upper wall speed $U=4.0\,\sigma/\tau$, the orientation
of more rigid chain segments along the shear flow direction produces
slightly higher density in the $3$rd, $4$th, and $5$th fluid layers
[\,see Fig.\,\ref{Fig:mol_dens_k}\,(b)\,].

\begin{figure}[t]
\includegraphics[width=12.cm,angle=0]{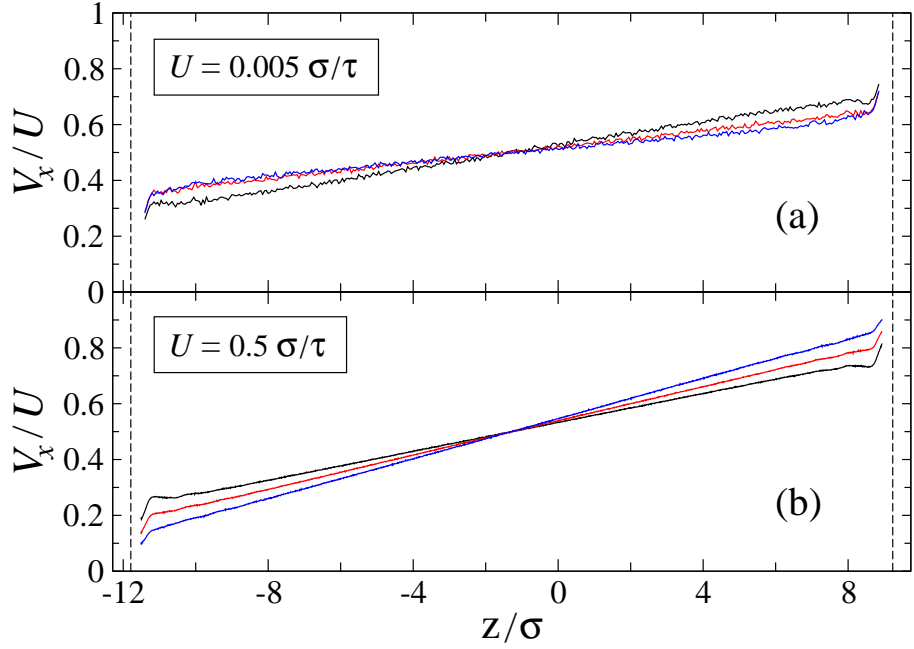}
\caption{(Color online) Averaged normalized velocity profiles for
the upper wall speeds (a) $U=0.005\,\sigma/\tau$ and (b)
$U=0.5\,\sigma/\tau$ and bending stiffness coefficients
$k_{\theta}=0.0\,\varepsilon$ (black lines),
$k_{\theta}=2.0\,\varepsilon$ (red lines), and
$k_{\theta}=3.0\,\varepsilon$ (blue lines).   The vertical axes
coincide with the location of the fcc lattice planes (at
$z/\sigma=-12.29$ and $9.73$).   The vertical dashed lines (at
$z/\sigma=-11.79$ and $9.23$) indicate reference planes for
computing the slip length.  } \label{Fig:velocity_prl}
\end{figure}

The representative velocity profiles are plotted in
Fig.\,\ref{Fig:velocity_prl} for the lowest $U=0.005\,\sigma/\tau$
and intermediate $U=0.5\,\sigma/\tau$ upper wall speeds and
$k_{\theta}=0.0\,\varepsilon$, $2.0\,\varepsilon$, and
$3.0\,\varepsilon$.  For $U=0.005\,\sigma/\tau$, despite extensive
averaging, the data remain noisy because the average flow velocity
is much smaller than the thermal fluid velocity $v_T=k_BT/m$.  In
all cases, the velocity profiles are anti-symmetric with respect to
the center of the channel and linear except within about $2\,\sigma$
near the walls.   Surprisingly, the dependence of slip velocity on
bending stiffness shows opposite trends for the reported upper wall
speeds; namely, the slip velocity for flexible chains is smaller for
$U=0.005\,\sigma/\tau$ in Fig.\,\ref{Fig:velocity_prl}\,(a), while
it is larger for $U=0.5\,\sigma/\tau$ in
Fig.\,\ref{Fig:velocity_prl}\,(b).  This result illustrates that the
effect of chain stiffness on the interfacial slip strongly depends
on flow conditions.  In what follows, the shear rate was extracted
from the linear part of velocity profiles excluding the interfacial
regions of about $4\,\sigma$.   As usual, the slip length was
computed by linear extrapolation of the velocity profiles to the
values $V_x=0$ below the lower wall and $V_x=U$ above the upper wall
and then averaged.

\subsection{Shear viscosity and slip length}
\label{subsec:viscosity_slip_length}

\begin{figure}[t]
\includegraphics[width=12.cm,angle=0]{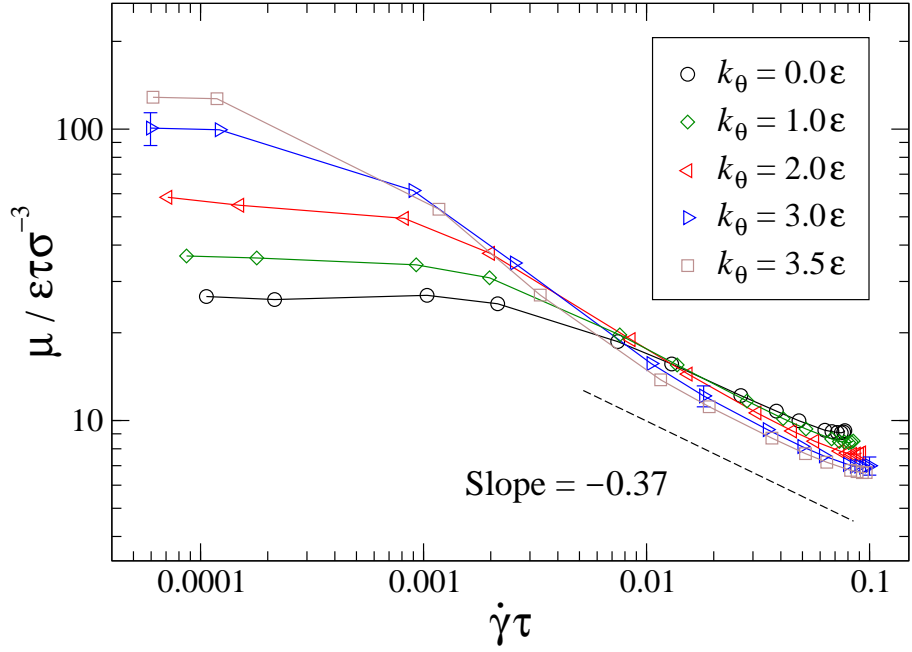}
\caption{(Color online) Shear rate dependence of the polymer
viscosity $\mu$ (in units of $\varepsilon\tau\sigma^{-3}$) for
selected values of the bending stiffness coefficient. The dashed
line indicates a slope of $-0.37$. Solid curves are a guide for the
eye.} \label{Fig:visc_shear_all}
\end{figure}

In steady shear flow, the fluid viscosity is defined by the relation
$\sigma_{xz}=\mu(\dot{\gamma})\,\dot{\gamma}$, where $\dot{\gamma}$
denotes the shear rate and $\sigma_{xz}$ is the shear stress through
any plane parallel to the solid walls.  In our simulations, the
shear stress per unit area was computed at the liquid-solid
interface by averaging the total force (in the shear flow direction)
between the lower wall atoms and the fluid molecules.  The
dependence of polymer viscosity on shear rate is presented in
Fig.\,\ref{Fig:visc_shear_all} for selected values of the bending
stiffness coefficient.   For more flexible chains
($k_{\theta}\leqslant2.0\,\varepsilon$), the gradual transition from
the Newtonian to shear-thinning regimes is clearly observed in the
accessible range of shear rates.  The dashed line with the slope
$-0.37$ is shown for reference in Fig.\,\ref{Fig:visc_shear_all},
indicating shear-thinning behavior of flexible chains
($k_{\theta}=0.0\,\varepsilon$ and $N=20$) reported in previous
studies~\cite{Priezjev08,Priezjev10}.   The characteristic shear
rate of the transition correlates well with the inverse relaxation
time of polymer chains in the bulk region [see inset in
Fig.\,\ref{Fig:auto_bulk_walls}\,(a)].  Not surprisingly, with
increasing chain stiffness, the polymer viscosity at low shear rates
increases, and the slope of the shear-thinning region becomes more
steep due to the orientation of partially uncoiled chains in the
shear flow.  The apparent saturation of the viscosity at high shear
rates is due to an increase in the fluid temperature near
interfaces.  The error bars are larger at low shear rates due to the
enhanced statistical uncertainty in averaging velocity profiles and
wall shear stress.

\begin{figure}[t]
\includegraphics[width=12.cm,angle=0]{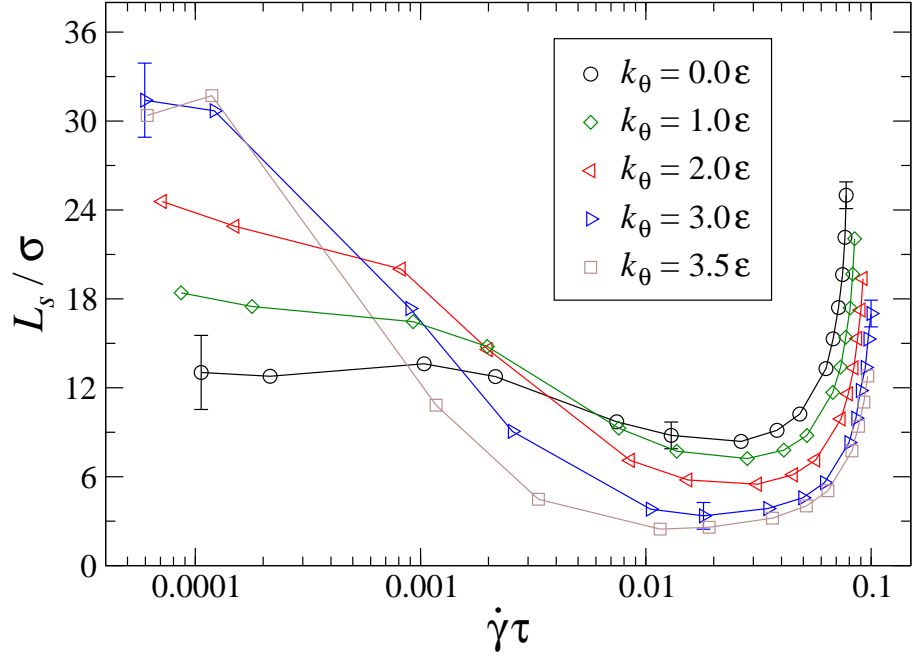}
\caption{(Color online) Variation of the slip length $L_s/\sigma$ as
a function of shear rate for the indicated values of the bending
stiffness coefficient. The solid curves are drawn to guide the eye.}
\label{Fig:shear_ls_all}
\end{figure}

Figure\,\ref{Fig:shear_ls_all} shows the dependence of slip length
as a function of shear rate for the same flow conditions and values
of the bending stiffness coefficient as in
Fig.\,\ref{Fig:visc_shear_all}.  All curves in
Fig.\,\ref{Fig:shear_ls_all} exhibit the same characteristic
feature: a pronounced minimum at low shear rates and a steep
increase at higher shear rates.  In case of flexible chains, this
behavior was analyzed
previously~\cite{Priezjev08,Priezjev09,Priezjev10} using
Eq.\,(\ref{Eq:ls_fr_k}).  As illustrated in
Fig.\,\ref{Fig:shear_ls_all}, the slip length (for
$k_{\theta}=0.0\,\varepsilon$) is nearly constant at low shear rates
because of the extended Newtonian regime [in
Fig.\,\ref{Fig:visc_shear_all}] and velocity-independent friction
coefficient.  With increasing shear rate, the relative competition
between the shear-thinning viscosity and the dynamic friction
coefficient in Eq.\,(\ref{Eq:ls_fr_k}) leads to a minimum in the
slip length, which is followed by a rapid increase at higher shear
rates.  Unexpectedly, increasing the chain stiffness produces larger
slip lengths at low shear rates but smaller $L_s$ at high shear
rates.  This trend can be understood by analyzing the effect of
chain stiffness on the friction coefficient as a function of the
slip velocity (see next subsection).

As mentioned previously, the range of the upper wall speeds
considered in the present study corresponds to anti-symmetric
velocity profiles so that the slip velocity is the same at the lower
and upper walls.  It is expected that at higher upper wall speeds,
the slip velocity at one of the solid walls will be much larger than
the fluid thermal velocity producing slip lengths much larger than
the channel height~\cite{LichPRL08}. The investigation of the slip
transition at very high shear rates is not the main focus of this
paper; and, therefore, it was not studied in detail.

\subsection{The dynamic friction coefficient}
\label{subsec:friction_velocity}

In this subsection, we analyze the influence of chain stiffness on
the friction coefficient at the liquid-solid interface as a function
of the slip velocity.  The results of previous MD studies have shown
that the data for flexible polymer chains and weakly attractive
crystalline surfaces can be well fitted by the following empirical
equation:
\begin{equation}
k/k^{\ast}=[1+(V_s/V_s^{\ast})^2]^{-0.35},
\label{Eq:friction_law}
\end{equation}
where the parameter $k^{\ast}$ is the friction coefficient at small
slip velocities and $V_s^{\ast}$ is the characteristic slip velocity
of the transition to the nonlinear
regime~\cite{Priezjev08,Priezjev09,Priezjev10}.  It was demonstrated
numerically that the friction coefficient $k^{\ast}$ is determined
by the contact density and the in-plane structure factor of the
first fluid layer~\cite{Priezjev10}.  Furthermore, the
characteristic slip velocity $V_s^{\ast}$ was found to correlate
well with the diffusion rate of fluid monomers over the distance
between nearest minima of the substrate potential~\cite{Priezjev10}.
The physical origin of the exponent $-0.35$ in
Eq.\,(\ref{Eq:friction_law}) is at present unclear.

Although the friction coefficient can be readily computed from
Eq.\,(\ref{Eq:ls_fr_k}), the slight curvature in the velocity
profiles near solid walls and the location of the liquid-solid
interfaces used to compute the slip length [\,see
Fig.\,\ref{Fig:velocity_prl}\,], introduce a small discrepancy
between the definitions $k(V_s)=\mu/L_s$ and
$k(V_{1})=\sigma_{xz}/V_{1}$, where $V_s=L_s\dot{\gamma}$ and
$V_{1}$ is the velocity of first fluid layer. To eliminate this
uncertainty, in the present study, the slip velocity was computed
directly from the velocity profiles as follows:
\begin{equation}
V_{1}=\int_{z_0}^{z_1}\!V_x(z)\rho(z)dz \,\Big/
\int_{z_0}^{z_1}\!\rho(z)dz, \label{Eq:velo_defin_eq}
\end{equation}
where the limits of integration ($z_0=-11.54\,\sigma$ and
$z_1=-10.87\,\sigma$) define the width of the first peak in density
profiles, which are shown for example in Fig.\,\ref{Fig:mol_dens_k}.

\begin{figure}[t]
\includegraphics[width=12.cm,angle=0]{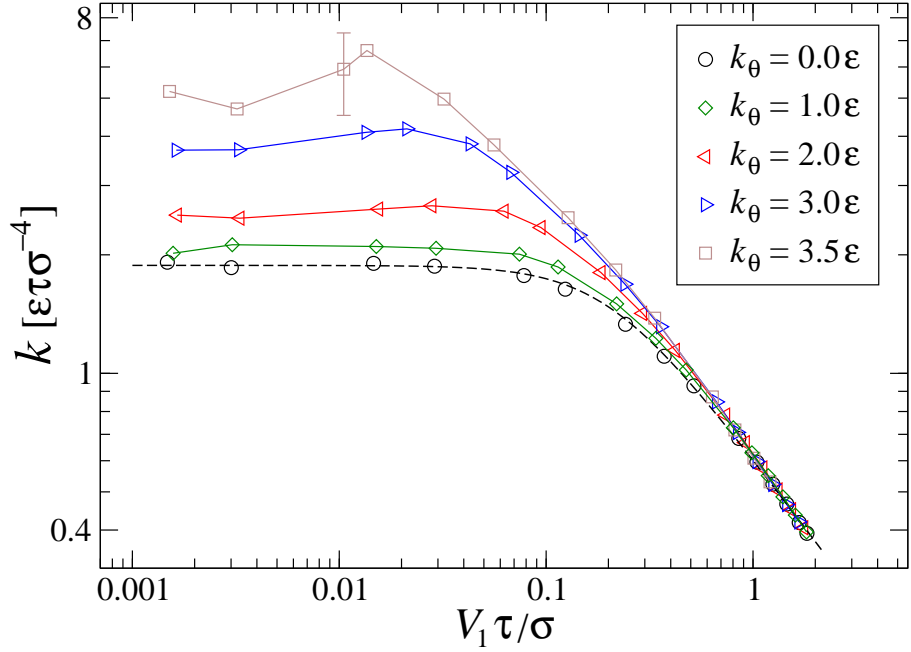}
\caption{(Color online) Log-log plot of the friction coefficient
$k=\sigma_{xz}/V_{1}$ (in units of $\varepsilon\tau\sigma^{-4}$) as
a function of the slip velocity of the first fluid layer $V_{1}$ (in
units of $\sigma/\tau$) for the tabulated values of the bending
stiffness coefficient. The dashed curve is the best fit to
Eq.\,(\ref{Eq:friction_law}) with
$k^{\ast}=1.88\,\varepsilon\tau\sigma^{-4}$ and
$V_s^{\ast}=0.2\,\sigma/\tau$. The solid curves are guides for the
eye. } \label{Fig:friction_velo}
\end{figure}

The friction coefficient $k(V_1)$ as a function of the slip velocity
is plotted in Fig.\,\ref{Fig:friction_velo} for several values of
the bending stiffness coefficient.  The important conclusion from
the present results is that, with increasing chain stiffness, the
friction coefficient at small slip velocities increases, and its
decay rate at large slip velocities is independent of the chain
stiffness.  It can be further observed that for more flexible
chains, $k_{\theta}\leqslant1.0\,\varepsilon$, the data are well
described by the functional form given by
Eq.\,(\ref{Eq:friction_law}).  However, as the chain stiffness
increases, the data in Fig.\,\ref{Fig:friction_velo} indicate
qualitative changes in the velocity dependence of the friction
coefficient, i.e., the appearance of a pronounced maximum at small
slip velocities.  This non-monotonic behavior is related to the
enhanced ordering of semiflexible chains near interfaces due to
their orientation along the shear flow direction. This effect will
be discussed in more detail in the next subsection.  For the largest
value of the stiffness coefficient, $k_{\theta}=3.5\,\varepsilon$,
the error bars are relatively large at small slip velocities because
the orientation of the extended chain segments in the first fluid
layer is strongly influenced by the sixfold symmetry of the wall
lattice and their relaxation dynamics is very slow [see
Fig.\,\ref{Fig:auto_bulk_walls}\,(b)].

Nevertheless, some trends in the nonlinear rate dependence of the
slip length presented in Fig.\,\ref{Fig:shear_ls_all} can be
understood from Eq.\,(\ref{Eq:ls_fr_k}) and the data reported in
Figs.\,\ref{Fig:visc_shear_all} and \ref{Fig:friction_velo}.  For
example, the ratio of shear viscosity to the friction coefficient is
smaller for stiffer chains with $k_{\theta}=3.5\,\varepsilon$ at
high shear rates, while the largest slip length at low shear rates
is reported for chains with $k_{\theta}=3.0\,\varepsilon$.  Also,
the sharp decay of the slip length for the cases
$k_{\theta}=3.0\,\varepsilon$ and $3.5\,\varepsilon$ in
Fig.\,\ref{Fig:shear_ls_all} is related to the large negative slope
of the polymer viscosity at low shear rates.  As mentioned earlier,
the nearly constant value of the slip length at low shear rates for
fully flexible chains in Fig.\,\ref{Fig:shear_ls_all} is due to the
Newtonian viscosity and a wide linear regime of friction determined
by the parameter $V_s^{\ast}$ in Eq.\,(\ref{Eq:friction_law}).

\subsection{Fluid structure near solid walls}
\label{subsec:friction_structure}

The examples of the fluid density profiles shown in
Fig.\,\ref{Fig:mol_dens_k} demonstrate that the fluid density
layering is most pronounced for the fluid monomers in contact with
wall atoms.  It is well known that, in addition to the fluid
ordering perpendicular to the substrate, the periodic surface
potential induces structure formation within the first fluid layer.
The measure of the induced order is the in-plane static structure
factor, which is defined as follows:
\begin{equation}
S(\mathbf{k})=\frac{1}{N_{\ell}}\,\,\Big|\sum_{j=1}^{N_{\ell}}
e^{i\,\mathbf{k}\cdot\mathbf{r}_j}\Big|^2,
\label{Eq:structure_factor}
\end{equation}
where the sum is over $N_{\ell}$ fluid monomers in the layer and
$\mathbf{r}_j=(x_j,y_j)$ is the position vector of the $j$-th
monomer.  Depending on the strength of wall-fluid interactions and
commensurability of liquid and solid structures, the structure
factor typically contains a set of sharp peaks at the reciprocal
lattice vectors, which are superimposed on several concentric rings
characteristic of the liquid-like short range
order~\cite{Thompson90}. In the past, several MD studies have
demonstrated a strong correlation between the magnitude of the
largest peak at the first reciprocal lattice vector and the friction
coefficient at liquid-solid
interfaces~\cite{Thompson90,Thompson95,Barrat99fd,Priezjev04,
Priezjev07,PriezjevJCP,Priezjev08,Priezjev09,Priezjev10}.

\begin{figure}[t]
\includegraphics[width=12.cm,angle=0]{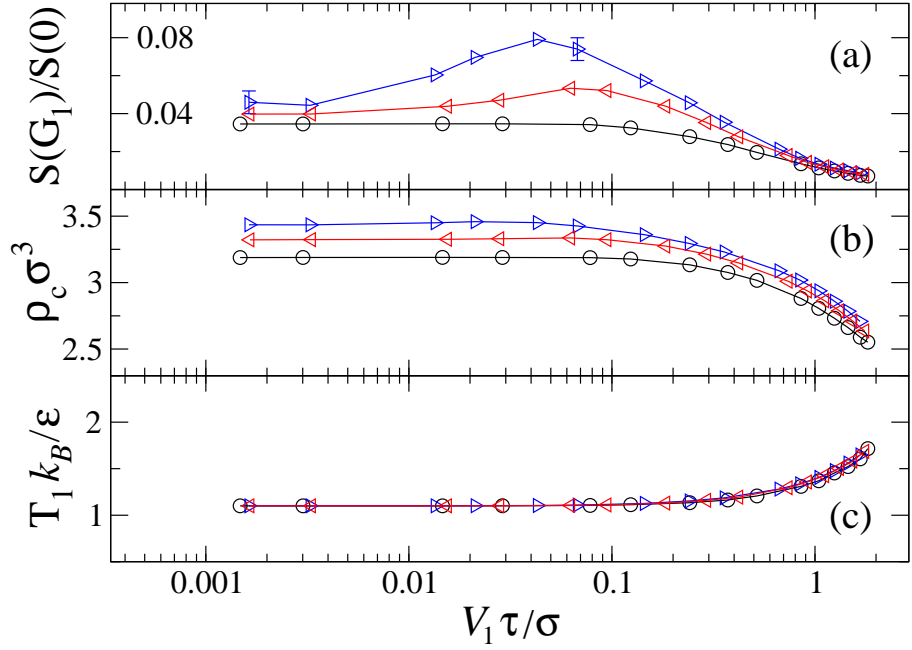}
\caption{(Color online) The normalized structure factor at the main
reciprocal lattice vector $\mathbf{G}_{1}=(7.23\,\sigma^{-1},0)$
(a), contact density (b), and temperature (c) of the first fluid
layer as a function of the slip velocity $V_{1}$ (in units of
$\sigma/\tau$). The values of the bending stiffness coefficient are
$k_{\theta}=0.0\,\varepsilon$ ($\circ$),
$k_{\theta}=2.0\,\varepsilon$ ($\triangleleft$), and
$k_{\theta}=3.0\,\varepsilon$ ($\triangleright$).}
\label{Fig:velo_k0_k2_k3_T_rho_sk}
\end{figure}

We next plot the dependence of the normalized structure factor
evaluated at the main reciprocal lattice vector
$\mathbf{G}_{1}=(7.23\,\sigma^{-1},0)$, the contact density, and the
temperature of the first fluid layer in
Fig.\,\ref{Fig:velo_k0_k2_k3_T_rho_sk} for three values of the
bending stiffness coefficient.  Similar to previous findings for
flexible chains~\cite{Priezjev09}, all three parameters in
Fig.\,\ref{Fig:velo_k0_k2_k3_T_rho_sk} remain constant at small slip
velocities, $V_1\lesssim0.1\,\sigma/\tau$, while the induced
structure [$S(\mathbf{G}_{1})/S(0)$ and $\rho_c$] reduces and the
fluid temperature increases at higher slip velocities.  In sharp
contrast, the structure factor for semiflexible chains exhibits a
distinctive maximum at small slip velocities.  Note that these
changes in the structure factor are not reflected in the contact
density, suggesting that they are mainly caused by the reorientation
of chain segments in the first layer along the shear flow direction.

\begin{figure}[t]
\includegraphics[width=12.cm,angle=0]{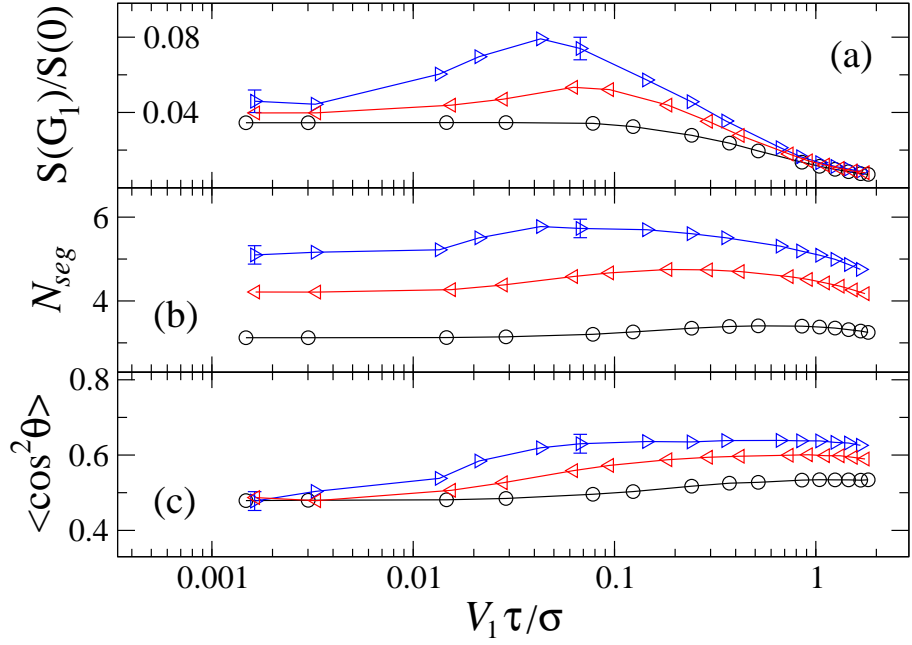}
\caption{(Color online) The structure factor (a), number of
consecutive monomers per chain (b), and bond orientation (c) in the
first fluid layer as a function of the slip velocity.  The values of
the bending stiffness coefficient are $k_{\theta}=0.0\,\varepsilon$
($\circ$), $k_{\theta}=2.0\,\varepsilon$ ($\triangleleft$), and
$k_{\theta}=3.0\,\varepsilon$ ($\triangleright$). The data for
$S(\mathbf{G}_{1})/S(0)$ are the same as in
Fig.\,\ref{Fig:velo_k0_k2_k3_T_rho_sk}\,(a). }
\label{Fig:velo_k0_k2_k3_S_Nseg_cos2}
\end{figure}

In order to quantify this hypothesis, we examined the chain
structure in contact with the substrate.
Figure\,\ref{Fig:velo_k0_k2_k3_S_Nseg_cos2} shows the average number
of consecutive monomers per chain in the first fluid layer and their
bond orientation with respect to the shear flow direction.
Specifically, we computed the average value
$\langle\textrm{cos}^2\theta\rangle$, where $\theta$ is the angle
between the $\hat{x}$ axis and the three-dimensional bond vector
connecting two consecutive monomers in the first fluid layer. In
this definition, $\langle\textrm{cos}^2\theta\rangle=0.5$ for the
planar isotropic distribution, whereas
$\langle\textrm{cos}^2\theta\rangle=1.0$ for the parallel
arrangement of bond vectors along the $\hat{x}$ axis.  The plots in
Fig.\,\ref{Fig:velo_k0_k2_k3_S_Nseg_cos2} reveal that, with
increasing chain stiffness, the first fluid layer consists of more
extended chain segments, which become preferentially aligned in the
direction of shear flow.  Notice that the orientation of flexible
chain segments remain isotropic at small slip velocities,
$V_1\lesssim0.1\,\sigma/\tau$.   Hence, the results in
Figs.\,\ref{Fig:velo_k0_k2_k3_T_rho_sk} and
\ref{Fig:velo_k0_k2_k3_S_Nseg_cos2} indicate that the appearance of
a maximum in $S(\mathbf{G}_{1})/S(0)$, which in turn affects the
friction coefficient in Fig.\,\ref{Fig:friction_velo}, is due to the
shear-induced alignment of semiflexible chain segments in the first
fluid layer.

\begin{figure}[t]
\includegraphics[width=12.cm,angle=0]{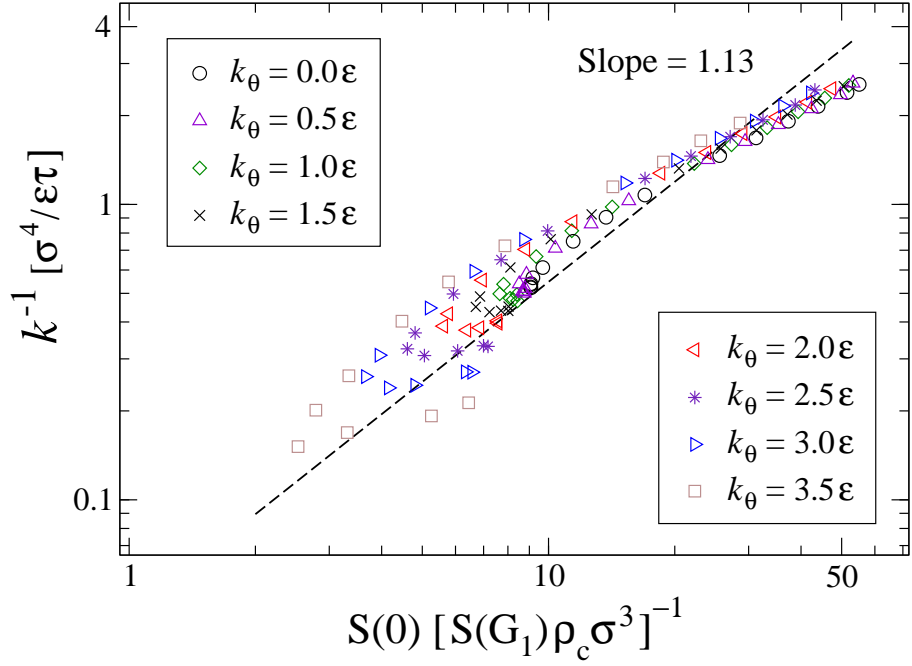}
\caption{(Color online) Log-log plot of the inverse friction
coefficient $k^{-1}=V_1/\sigma_{xz}$ (in units of
$\sigma^4/\varepsilon\tau$) as a function of
$S(0)/\,[S(\mathbf{G}_1)\,\rho_c\,\sigma^3]$ computed in the first
fluid layer.  The values of the bending stiffness coefficient are
tabulated in the inset.  The dashed line $y=0.041\,x^{1.13}$ is
taken from~\cite{Priezjev10} and it is shown for reference. }
\label{Fig:inv_fr_vs_S0_div_S7_ro_c_low}
\end{figure}

We finally summarize our data by plotting the inverse friction
coefficient as a function of the combined variable
$S(0)/[S(\mathbf{G}_1)\,\rho_c]$ in
Fig.\,\ref{Fig:inv_fr_vs_S0_div_S7_ro_c_low}.   It was previously
shown for flexible polymer chains that in the linear regime
[$V_s<V_s^{\ast}$ in Eq.\,(\ref{Eq:friction_law})], the friction
coefficient can be described by a function of the variable
$S(0)/[S(\mathbf{G}_1)\,\rho_c]$ for a number of material parameters
of the interface, such as fluid and wall densities, surface energy,
chain length, and wall lattice type~\cite{Priezjev10}.  The best fit
to the MD data taken from~\cite{Priezjev10} is indicated by the
dashed line in Fig.\,\ref{Fig:inv_fr_vs_S0_div_S7_ro_c_low}.   It
can be observed that, for all values of the bending stiffness
coefficient, the data points are distributed around the straight
dashed line.  In agreement with the previous
results~\cite{Priezjev10}, at small slip velocities, the friction
coefficient for more flexible chains,
$k_{\theta}\leqslant2.0\,\varepsilon$, is well described by the
master curve. The most noticeable difference between flexible and
semiflexible chains in Fig.\,\ref{Fig:inv_fr_vs_S0_div_S7_ro_c_low}
is the appearance of the hook-shaped curvature at small slip
velocities, which is related to the local maximum in the structure
factor discussed earlier.  In other words, the same value of the
product of structure factor and contact density,
$S(0)/[S(\mathbf{G}_1)\,\rho_c]$, corresponds to two different
values of the friction coefficient, depending on the slip velocity
and chain stiffness.

In summary, the results in
Fig.\,\ref{Fig:inv_fr_vs_S0_div_S7_ro_c_low} for semiflexible
chains, $k_{\theta}\leqslant2.0\,\varepsilon$, confirm previous
findings that the friction coefficient at small slip velocities is
determined by the magnitude of the surface-induced peak in the
structure factor and the contact density of the first fluid
layer~\cite{Priezjev10}.  The deviation from the master curve for
more rigid chains, $k_{\theta}>2.0\,\varepsilon$, might be related
to the slower relaxation dynamics of the chains in the interfacial
region, similar to the trends found in dense polymer films at low
shear rates~\cite{Priezjev09}.

\section{Conclusions}

In this paper, we have presented results from extensive molecular
dynamics simulations of thin polymer films confined by crystalline
walls with weak surface energy.  The computations were based on a
coarse-grained bead-spring model of linear polymer chains with an
additional bond angle potential that controls chain bending
stiffness.  The spatial configuration and local relaxation dynamics
of polymer chains were characterized by the radius of gyration and
the decay rate of the autocorrelation function of the first normal
mode.  We found that semiflexible chains near solid walls become
more uncoiled and their relaxation dynamics is significantly slowed
down.

The most interesting result of the present study is the appearance
of a distinct maximum in the velocity dependence of the friction
coefficient due to the shear-induced alignment of semiflexible chain
segments in the first fluid layer near solid walls.  At small slip
velocities, the orientation of more extended chain segments along
the flow direction produces an enhanced ordering within the first
fluid layer measured by the height of the main peak in the structure
factor.  This effect is absent for fully flexible chains since their
segment orientation in the adjacent layer remains isotropic at small
slip velocities.  In addition, it was demonstrated that, with
increasing slip velocity, the decay of the friction coefficient is
independent of the chain stiffness.

Our simulation results indicate that the main features in the shear
rate dependence of the slip length include a nearly constant value
at low shear rates, a pronounced minimum at intermediate rates, and
a rapid increase at high shear rates.  These slip flow regimes are
determined by the ratio of the rate-dependent polymer viscosity and
the dynamic friction coefficient.  Overall, we conclude that it is
difficult to predict the net effect of chain stiffness on the slip
length without performing numerical simulations; especially at low
shear rates, where both polymer viscosity and friction coefficient
increase with increasing bending rigidity.

\section*{Acknowledgments}

Financial support from the National Science Foundation
(CBET-1033662) is gratefully acknowledged. Computational work in
support of this research was performed at Michigan State
University's High Performance Computing Facility.


\bibliographystyle{prsty}

\end{document}